\documentclass[preprint]{elsarticle}
\usepackage{amsmath,amsthm,amsfonts,graphicx,textcomp,dcolumn,bm,hyperref}
\usepackage{float}
\usepackage[mathlines]{lineno}
\usepackage[USenglish]{babel}
\usepackage[table]{xcolor}

\journal{Carbon}

\begin{document}

\begin{frontmatter}
\title{Observation of Coulomb blockade in nanostructured epitaxial bilayer graphene on SiC}

\author[cav]{Cassandra~Chua\corref{cor1}}
\ead{cassandra.chua@cantab.net}
\author[cha]{Arseniy~Lartsev}
\author[cav]{Jinggao~Sui}
\author[npl]{Vishal~Panchal}
\author[cav]{Reuben~Puddy}
\author[cav]{Carly~Richardson}
\author[cav]{Charles~G.~Smith}
\author[npl]{T.J.B.M.~Janssen}
\author[npl,rhl]{Alexander~Tzalenchuk}
\author[lkp]{Rositsa~Yakimova}
\author[cha]{Sergey~Kubatkin}
\author[cav]{Malcolm~R.~Connolly\corref{cor1}}
\ead{mrc61@cam.ac.uk}

\cortext[cor1]{Corresponding Author}

\address[cav]{Cavendish Laboratory, 19 JJ Thomson Avenue, Cambridge CB3 0HE, United Kingdom}
\address[cha]{Department of Microtechnology and Nanoscience, Chalmers University of Technology, S-412 96 G\"{o}teborg, Sweden}
\address[npl]{National Physical Laboratory, Hampton Road, Teddington, TW11 0LW, United Kingdom}
\address[rhl]{Royal Holloway, University of London, Egham, TW20 0EX, United Kingdom}
\address[lkp]{Department of Physics, Chemistry and Biology (IFM), Link\"{o}ping University, S-581 83 Link\"{o}ping, Sweden}

\begin{abstract}

We study electron transport in nanostructures patterned in bilayer graphene patches grown epitaxially on SiC as a function of doping, magnetic field, and temperature. Away from charge neutrality transport is only weakly modulated by changes in carrier concentration induced by a local side-gate. At low n-type doping close to charge neutrality, electron transport resembles that in exfoliated graphene nanoribbons and is well described by tunnelling of single electrons through a network of Coulomb-blockaded islands. Under the influence of an external magnetic field, Coulomb blockade resonances fluctuate around an average energy and the gap shrinks as a function of magnetic field. At charge neutrality, however, conduction is less insensitive to external magnetic fields. In this regime we also observe a stronger suppression of the conductance below $T^*$, which we interpret as a sign of broken interlayer symmetry or strong fluctuations in the edge/potential disorder.

\end{abstract}


\end{frontmatter}

\section{Introduction}
Graphene layers grown epitaxially on SiC wafers are an attractive solution for upscaling graphene-based electronic devices for a variety of applications such as sensing, spintronics, and electrical metrology \cite{ber04, ber06, deh07, Beshkova2016}. Although SiC provides a naturally insulating substrate and direct growth avoids contamination and sample degradation incurred during transfer or exfoliation, reproducing the behaviour of pristine exfoliated prototypes is not straightforward due to patches of bilayer graphene \cite{yag15} and interactions with the underlying SiC substrate. Charge transfer from interfacial states in the buffer layer actually improves the robustness of graphene for quantum resistance metrology and provides a natural mechanism for breaking layer-symmetry and opening a band gap in bilayers \cite{mcc06b}, but may have an adverse affect on carrier mobility and tunability. The development of single-electron tunnelling spectroscopy in SiC graphene would not only provide additional insights about the graphene-SiC interacation but also enable the fabrication of large arrays of single-electron quantum devices such as pumps \cite{con13} and spin qubits \cite{Trauzettel2008}. SiC graphene is particularly attractive as it is inherently scalable, does not need to be transferred, and can potentially be integrated with silicon. Unlike exfoliated graphene nanoribbons (GNRs), which exhibit sharp conductance peaks arising from resonant Coulomb-blockade transmission between electron-hole puddle and edge-induced quantum dots around the Dirac point \cite{han07,Bischoff2015}, single-electron charing effects have so far eluded detection in  graphene on SiC. The absence of this behaviour was attributed to the weaker disorder potential \cite{hwa15}. In addition, the strong $n$-type doping of as-grown material \cite{kop10, tza10,jan11} must be neutralised in order to tune to the Dirac point\cite{li_09,pan14} where tunnelling and single-electron charging effects can be observed. Methods have been developed for tuning the carrier density, such as decoupling the graphene from the substrate \cite{vir10,pal14,lin15,oli15}, top-gating\cite{li_09,she12}, photochemical gating\cite{lar11s}, or corona discharge gating \cite{lar14a,hua15}, but have yet to be used in single-electron devices. 

\section{Sample Preparation and Measurement}
In this paper we use a combination of side- and corona-discharge gating to tune the doping and report single-electron charging effects in nanostructured bilayer graphene patches on the Si-face (0001) of SiC substrates \cite{vir08}. Previous studies of exfoliated bilayer GNRs showed observable contributions from the bilayer band gap when the gap size was larger or comparable with the disorder potential (see Refs. \cite{wan13b} and \cite{Bischoff2015} and references therein), giving us a way to probe the interplay between energy gaps associated with broken inversion symmetry \cite{yu13}, one-dimensional confinement, and electron-electron interactions. Our devices were defined with electron-beam lithography and dry etched using an O$_2$ plasma. To modify the global carrier density we use corona-discharge gating \cite{lar14a}, which involves spraying charge on a dielectric layer spin-coated over the device, and a local graphene side gate for fine tuning the doping over a narrower range. Transport measurements were performed with fields up to $B$=8 T and temperatures down to $T$=1.4 K. We focus on the behaviour of a bilayer GNR device with width (W)$\approx$100 nm and length (L) of $\approx$700 nm with a side gate $\approx$180 nm from the device. A Kelvin probe micrograph \cite{Panchal2013} confirming the nanoribbon is bilayer graphene is shown in Fig. \ref{fig:LGNR}(a). Figure \ref{fig:LGNR}(b) shows a plot of the two-terminal conductance as a function of the number of negative discharge from the ion gun. As expected, the conductance drops due to the reduction in electron carrier density. Hall effect measurements from samples fabricated from similar wafers would suggest the doping of as-prepared devices is $~10^{13}$ carriers per cm$^2$. To examine the low-temperature behaviour at different carrier densities we measured at three stages of discharge doping and cooldowns. In the absence of precise knowledge of the doping we refer to these as high (HD), medium (MD), and low (LD) doping, indicated in Fig. \ref{fig:LGNR}(b).  

\section{Results and Discussion}
\begin{figure}[!t]
\centering
\includegraphics[width=85mm]{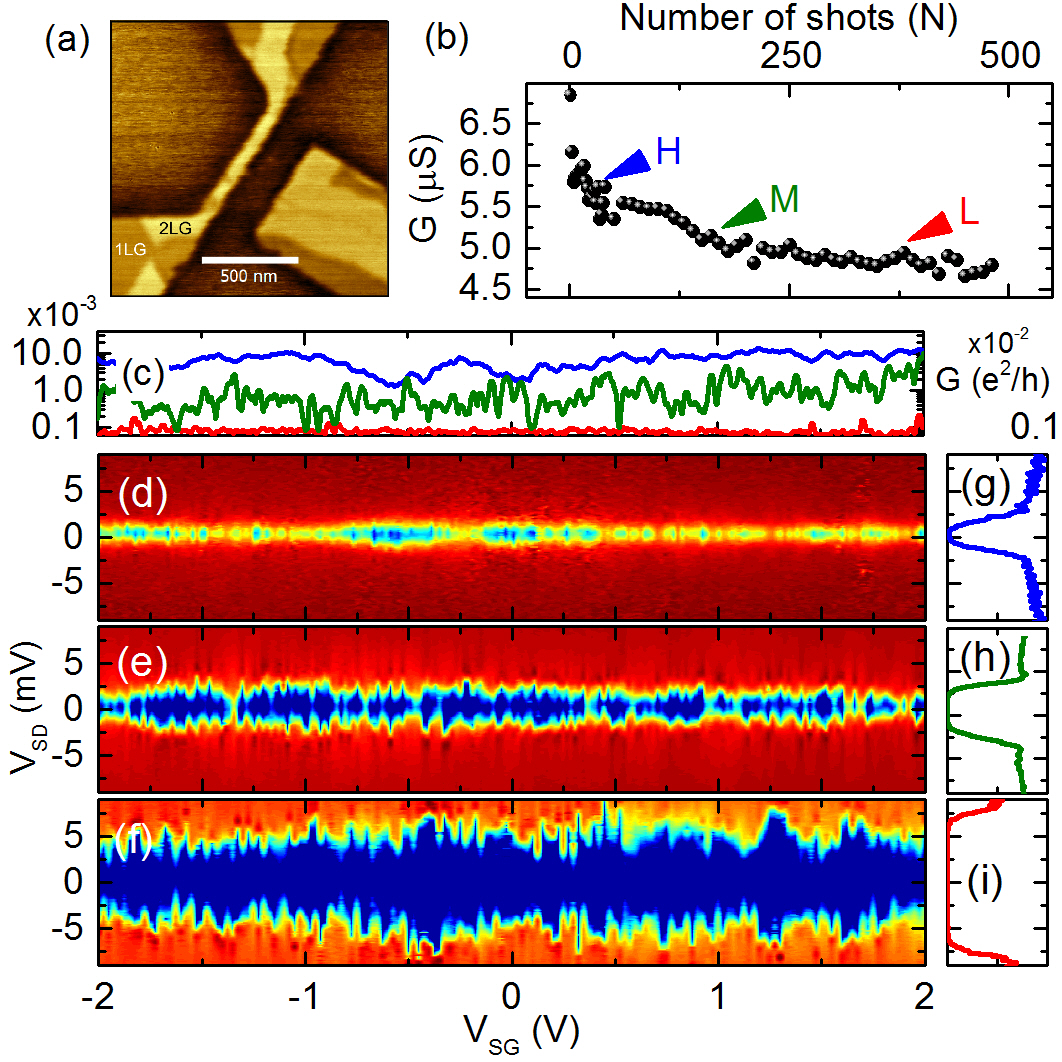}
\caption{(a) Kelvin probe micrograph of the graphene nanoribbon, where light and dark correspond to bilayer and single-layer graphene, respectively. (b) Conductance at room temperature as a function of the number of corona discharge shots, with arrows indicating the three different carrier concentrations studied. (c) Conductance as a function of side-gate voltage at HD (blue), MD (green), and LD (red). Conductance as a function of source drain bias (V$_{SD}$) and side gate voltage (V$_{SG}$) for (d) HD, (e) MD, and (f) LD. (g)-(i) Conductance as a function of source- drain bias showing the maximum gap. (T $\approx 1.4$ K).}
\label{fig:LGNR}
\end{figure}

Figure \ref{fig:LGNR}(c) shows a comparison of the linear conductance measured at $V_{SD}$=1 mV as a function of side-gate voltage ($V_{SG}$) at 1.4 K for each doping level. At HD and MD the conductance exhibits reproducible fluctuations, but for LD it is mostly within the noise floor. To uncover the origin of this behaviour we performed bias spectroscopy by sweeping V$_{SD}$ and V$_{SG}$ and plot the charge-stability diagrams in Fig. \ref{fig:LGNR}(d)-(f). At HD the conductance is suppressed but remains non-zero (soft gap) for $\approx$1 mV around $V_{SD}$=0 V, while at MD and LD $G$ exhibits a hard gap (where $G$ remains zero for a finite range of $V_{SD}$) of $\approx$2 mV and $\approx$5 mV, respectively [Fig. \ref{fig:LGNR}(g)-(i)]. Furthermore, the latter exhibit diamond or shard-like features reminiscent of the charge stability in a network of Coulomb-blockaded quantum dots. At LD the shards are less well defined with a periodicity in $V_{SG}$ of about 60 mV, but as they rarely close the total conductance at low bias is strongly suppressed for all $V_{SG}$.

\begin{figure}[!t]
\centering
\includegraphics{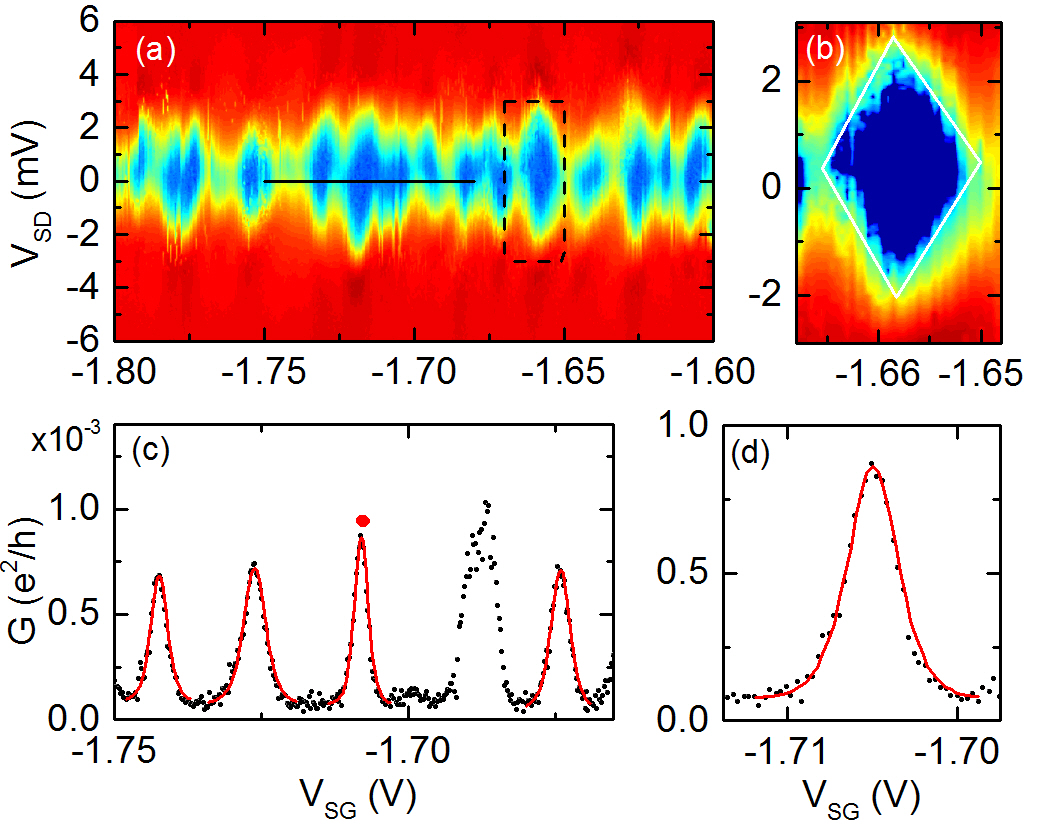}
\caption{(a) Conductance as a function of V$_{SD}$ and V$_{SG}$ at MD. (b) Representative Coulomb diamond outlined by the dashed box in (a). (c) Conductance as a function of V$_{SG}$ at zero bias along the line in (a), showing conductance resonances (black points) and fits (red lines) based on Coulomb blockade theory. (d) Detailed plot of the fitted peak indicated by a red dot in (c). (T $\approx 1.4$K).}
\label{fig:mdiamond}
\end{figure}

Concentrating first on the MD regime, Fig. \ref{fig:mdiamond}(a) shows a region of the charge stability diagram in more detail. We observe $\approx$10 diamonds with roughly uniform height and width. A typical diamond is shown in Fig. \ref{fig:mdiamond}(b) and a sweep at zero bias shows the periodic conductance resonances as a function of V$_{SG}$ [Fig. \ref{fig:mdiamond}(c)]. The maximum source-drain gap is $\approx$2 mV and consecutive resonances are spaced by $\Delta V_{SG}\approx$15 mV. The periodicity over this range strongly suggests that transport occurs either via a single or a few similar quantum dots with typical charging energy $E_c\approx$2 meV. To test this we fit the following equation describing single-electron tunnelling through many nearly degenerate states in the classical Coulomb blockade regime \cite{gol08}, $G = G_P(X/\sinh{[X]})$, where $G$ is the temperature-dependent conductance, $X$=$(\mu - E_0)/k_BT)$, $\mu$ is the chemical potential, $E_0$ is the energy of the resonant bound state where tunnelling occurs, $\Gamma = \Gamma_L + \Gamma_R$ ($\Gamma_L$ and $\Gamma_R$ are the tunnelling rates through the left and right barriers, respectively), and $G_P = (e^2/h)(\rho \Gamma_{L} \Gamma_{R}/(2 \Gamma)$ ($\rho$ is the density of bound states at the chemical potential $\mu$). Using $T$=1.4 K, and the relationship $E_0 = \alpha V_{SG}$, we obtain values for $G_P$ and $\alpha$ that can be compared with the lever arm deduced from the slope of their respective Coulomb diamonds. One of the fits is shown in figure~\ref{fig:mdiamond}(d) and we find reasonable general agreement between the measured and fitted lever arms, yielding an average $\alpha$ of $\approx$0.175 meV/V (See supplemental material \footnote{See supplemental material at [] for table of parameters for Coulomb-blockade resonances and data from a second device.} for measured parameters of each peak). 

\begin{figure}[!h]
\centering
\includegraphics{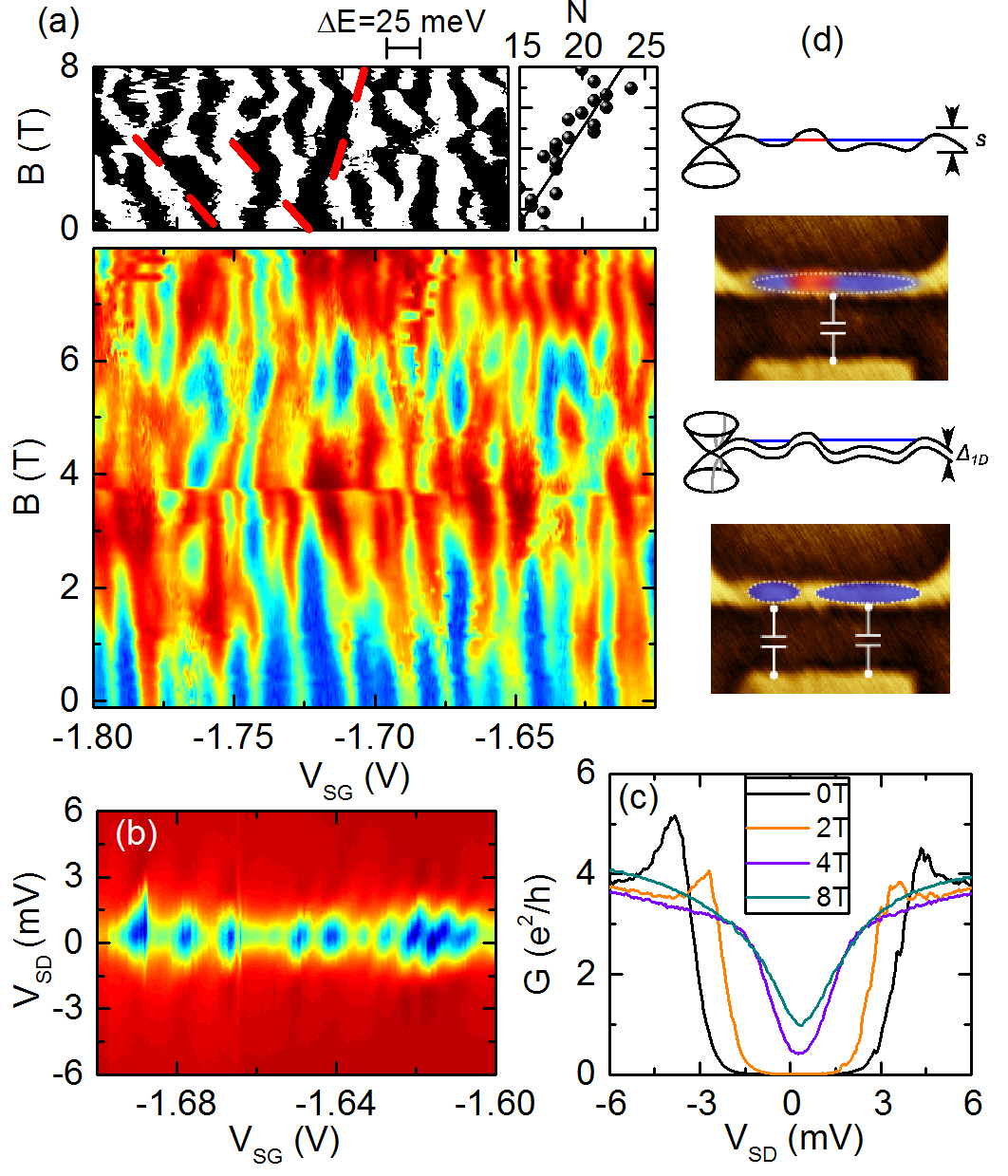}
\caption{(a) Conductance as a function of $V_{SG}$ and $B$ at MD (main panel). Raw data differentiated and segmented such that positive and negative slopes are white and black, respectively. Trajectories of peaks are highlighted by red lines (upper left). The number of peaks as a function of magnetic field is shown in upper right panel. (b) Conductance as a function of $V_{SD}$ and $V_{SG}$ at 8 T. (c) Plot of $G$ as a function $V_{SD}$ averaged across side gate voltage at 0, 2, 4 and 8 T. (d) Schematic diagrams showing a possible realisation of the disorder potential and quantum confinement gap along the GNR at $B=$0 T (lower) and $B=$8 T (upper). (T $\approx 1.4$K).} 
\label{fig:mmagneto}
\end{figure}

\begin{figure}[!h]
\centering
\includegraphics{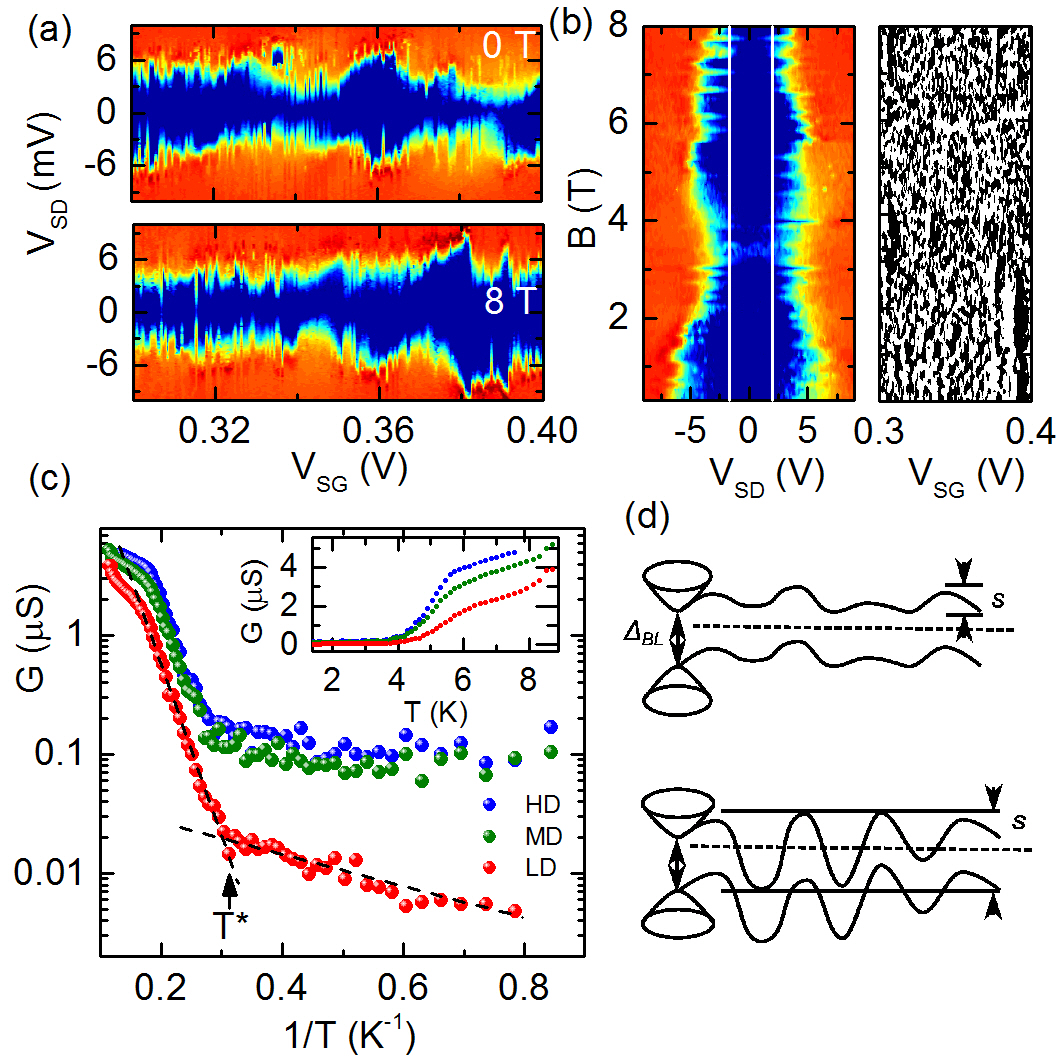}
\caption{(a) $G$ as a function of V$_{SD}$ and V$_{SG}$ at 0 T and 8 T at LD. (b) (Left) Gap in $G(V_{SD})$ measured at fixed side-gate voltage as a function of magnetic field. (Right) Side-gate sweeps as a function of magnetic field, differentiated and segmented in the same way as Fig. \ref{fig:mmagneto}(a), (T $\approx 1.4$K). (c) Dependence of the conductance as a function of inverse temperature for each doping level (raw data shown in the inset.) (d) Schematic energy as a function of position along the GNR for the case $s<\Delta_{BL}$ (upper), where transport occurs via variable-range hopping, and for thermal activation below $T^*$ if $s>\Delta_{BL}$ (lower).}
\label{fig:ldiamag}
\end{figure}

To explore the QD structure in more detail, Fig. \ref{fig:mmagneto}(a) shows the effect of a perpendicular magnetic field on the single-particle addition spectrum at MD. As a function of $B$, the Coulomb blockade resonances fluctuate around an average energy. This behavior is now well understood and arises from anticrossings between the single-particle levels \cite{gut09a,Libisch2010}. At a large particle number there are a large number of such anticrossings, so the resonances exhibit kinks and slopes $\approx$10 meV/T [Fig. \ref{fig:mmagneto}(a)] without shifting uniformly in energy \cite{Chiu2012}. The number of peaks also roughly doubles and the average period in $V_{SG}$ halves to around $5$ mV. At 8 T the gap in source-drain bias shrinks leading to smaller Coulomb diamonds [Fig. \ref{fig:mmagneto}(b)] and increase in the average conductance. Fig. \ref{fig:mmagneto}(c) plots the bias sweeps averaged across side-gate voltage at increasing magnetic fields and shows a softening of the gap for fields $>$4 T and a shrinking of the the gap to $\approx$1 meV by 8 T. This trend is also directly visible in the side-gate sweeps as a function of magnetic field shown in Fig. \ref{fig:mmagneto}(a). Such strong positive magnetoconductance is characteristic of GNRs and is associated with closing of both the transport and source-drain gaps \cite{Bai2010a}. Drawing from similar behavior in strongly disordered quasi-one-dimensional GaAs channels, this was explained by an increase in the characteristic size $L_c$ of the QDs and the consequent decrease in the energy required to hop between them. In graphene GNRs there is strong evidence that QDs form due to potential fluctuations, which in SiC have been described using Gaussian statistics parameterised by the strength $s\approx$10 meV, a factor of 5 less than on SiO$_2$. QDs form at MD when $E_F$ is within $s$, and tunnel barriers form between adjacent electron-hole puddles due to the quantum confinement gap $\Delta_{QC}$. For a 100 nm-wide GNR $\Delta_{QC}= \pi^2\hbar^2/m^*w^2\approx s$, the situation depicted in Fig. \ref{fig:mmagneto}(d). Transport is dominated by a few QDs and the conductance exhibits periodic transmission resonances [Fig. \ref{fig:mdiamond}(c)]. When a magnetic field is applied, the single-particle energy spectrum in bilayer graphene becomes $E_N=\pm\hbar\omega_c\sqrt{N(N-1)}$, where $\hbar$ is the reduced Planck's constant, $\omega_c=eB/m^*$ is the cyclotron frequency, $e$ is the electron charge, and $N$ is the orbital quantum number. A highly degenerate Landau level comprising states $N$=0,1 forms along $pn$ junctions where $E_F\approx$0. In this regime, the density of states decreases in the bulk of the puddles while it increases at their edges. Electron transport through the GNR is therefore not confined to a particular puddle but can be delocalized in the GNR owing to chiral edge channels~\cite{Chiu2012}. The consequent increase in the size $L_c$ of the islands and reduction in charging energy provides a natural explanation for the shorter period and smaller source-drain bias gap in Fig.~\ref{fig:mmagneto}, although we can not rule out other effects such as a field-dependent lever arm. $P-n$ junctions that form due to carrier inversion at the edges of the graphene may also enhance this short-circuiting effect in SiC-supported GNRs \cite{Panchal2013}. 
While the behaviour at MD fits within the original framework developed for potential/edge disorder-induced QDs in exfoliated monolayer GNRs \cite{Bischoff2015}, at LD the interpretation is more complicated due to presence of a vertical-field induced bandgap. It has been shown previously that the combined influence of charge transfer from a polymer and the buffer layer can open a gap of $\Delta_{BL}\approx$30 meV in bilayer patches on SiC \cite{chu14}. As implied already by the absence of peaks in the conductance at LD, resonant transmission through states at the band edges would be avoided since $s<\Delta_{BL}$ (See supplemental material \footnotemark[\value{footnote}] for similar data taken from a second device). The magnetotransport at LD is shown in Fig. \ref{fig:ldiamag} and is consistent with this: $\Delta V_{SD}$ does not shrink between 0 T to 8 T [see Fig. \ref{fig:ldiamag}(a)], and changes non-monotonically as a function of increasing magnetic field [Fig. \ref{fig:ldiamag}(b)]. Above $V_{SD}>$2 mV the shard-like features do appear as a function of $V_{SD}$, indicating that transport still proceeds via localized states above this energy, but the presence of such states complicates unambiguous extraction of $\Delta_{BL}$ from bias spectroscopy. 

Another way to probe gap formation is via the $T$-dependence of the conductance shown in Fig. \ref{fig:ldiamag}(c). To obtain this data we perform $G(V_{SG})$ sweeps at low temperature and fix $V_{SG}$ between resonances when changing temperature. Theoretically, three types of $T$-dependence have been previously invoked: at high temperature, thermal activation of electrons described by $G\propto\exp({E_a/2k_BT})$, either between adjacent localized states ($E_a=E_c$) or via extended states above a uniformly gapped region, $E_a=\Delta_{BL}$ or $\Delta_{QC}$. At low temperature, variable-range hopping (VRH) leads to $G\propto\exp(-({T_0/T})^{\gamma})$, where $T_0$ is the characteristic temperature for hopping, and $\gamma = 1/2$ for both 1D Mott and Efros-Shlovski VRH \cite{shk84}. We do observe a very clear change in behaviour at $T^*\approx$3.2 K for LD and 3.6 K for MD and HD, with $E_a\approx$6 meV for $T>T^*$ [Fig. \ref{fig:ldiamag}(c)]. VRH is expected to dominate when the thermal energy drops to roughly $1/10$ of the activation energy of an individual island and our measured $E_a/kT^*\approx$5 is in reasonable agreement. The similarity in the values of $E_a$ suggests that it is set by the (fixed) quantum confinement gap, rather than by $\Delta_{BL}$, which is larger close to the Dirac Point.

One possible feature of Fig. \ref{fig:ldiamag}(c) that betrays the influence of a bandgap at LD is the stronger suppression of $G$ below $T^*$. A linear fit based on thermal activation [Fig. \ref{fig:ldiamag}(c)], however, yields $E_a\approx$120 $\mu$eV, much smaller than the $\Delta_{BL}\approx$30 meV assumed earlier to account for the absence of transport resonances. Another explanation for the behaviour for LD below $T^*$ invokes the increased density of QDs at charge neutrality. The absence of conductance resonances in Fig. \ref{fig:LGNR}(c) is then naturally explained by stochastic Coulomb blockade and the greater improbability of simultaneously aligning energy levels in multiple quantum dots. 

\section{Conclusion}
In summary, we have studied electron transport in nanostructured SiC epitaxial bilayer graphene nanostructures as a function of doping, magnetic field, and temperature. The insulating state at low-temperature and away from charge neutrality exhibits sharp resonances and is well described within the framework for quantum dot formation in exfoliated GNRs, with multiple quantum dots forming in series due to the interplay between disorder and quantum confinement. Around the charge neutrality point conduction resonances are absent and transport is suppressed even at high magnetic field, consistent with a bandgap induced by broken layer symmetry. 

We would like to acknowledge support from Engineering and Physical Sciences Research Council (EP/L020963/1) and the EU project 696656 - GrapheneCore 1. Additional data related to this paper are available at the University of Cambridge data repository.

\section*{References}
\bibliography{SiCQDs}

\end{document}